\documentstyle[psfig,aps]{revtex}

\begin{document}

\twocolumn[\hsize\textwidth\columnwidth\hsize\csname@twocolumnfalse\endcsname

\title{Hole-pair symmetry and excitations in the strong-coupling extended
$t-J_z$ model}

\author{R.M. Fye}

\address{ Department of Physics and Astronomy,
University of New Mexico,
Albuquerque, NM 87131}


\maketitle

\begin{abstract}
We analytically calculate the ground state
pairing symmetry and excitation
spectra of two holes doped into the half-filled
$t- t^\prime -t^{\prime \prime}-J_z $ model in the strong-coupling
limit ($J_z >> |t|, |t^\prime|, |t^{\prime \prime}|$). For
the $t^\prime -t^{\prime \prime}-J_z $ model, there are
regions of $d$-wave,
$s$-wave, and $p$-wave symmetry.
We find that the 
$t-J_z$ model maps in lowest order onto the 
$t^\prime -t^{\prime \prime}-J_z $ model on the boundary between
$d$ and $p$ symmetry, with a flat lower branch of the
pair excitation spectrum. In higher order $d$-wave symmetry
is selected; however, we predict that
the addition of the appropriate
$t'$ and/or $t''$ should drive the hole-pair symmetry to $p$-wave.
We perturbatively construct an extended quasi-pair for the $t-J_z$ model.
We compare with analytic
calculations for a 2x2 plaquette and numerical
work, and discuss implications
for the experimentally relevant parameter regime.

\end{abstract}


]

Although the issue has not been completely
resolved, a variety of
experiments has indicated that the pair symmetry 
in the hole-doped
cuprate superconductors is either pure $d_{x^2 - y^2}$ or has a strong
$d_{x^2 - y^2}$ component\cite{dexp1,annett}. Theoretical and
numerical studies of the two-dimensional Hubbard, $t-J$, and related
models believed relevant to the high-$T_c$ compounds
have also suggested $d_{x^2 - y^2}$ 
pairing\cite{dagottorev,schrieffer,reviews,scalrev},
and Hubbard and $t-J$ models
on 2x2 plaquettes have recently provided an
intuitive picture of how such
a pair symmetry might arise\cite{scaltrug,whitescal}.
However, there are few rigorous theoretical results in this general
area.

Recent experimental work has indicated that a pseudogap
with the same symmetry as the superconducting gap
can persist above $T_c$
in underdoped cuprate
superconductors \cite{dexp1,pseudo1,pseudo2,preform}.
This, along with the short high-$T_c$ coherence
length \cite{dagottorev},
is generally consistent with a strong-coupling picture,
where pairs can preform at $T > T_c$ \cite{nan}.
Numerical work has in addition
suggested that the $t-J$ and $t-J_z$ models
have many similar properties\cite{dagottorev,dagottop,tjz}, and that 
the $t-J_z$ model may hence provide a suitable starting point for
understanding $t-J$ behavior\cite{cumulant}.
In this Letter, we consider two holes doped into the half-filled 
$t- t^\prime -t^{\prime \prime}-J_z$ model in the strong coupling limit
($J_z >> |t|, |t^\prime|, |t^{\prime \prime}|$).
We calculate the symmetry of the hole pair in the ground state
as well as the pair excitation spectrum.
We consider first the $t^\prime-J_z$ model,
and show how singlet pairs
can be constructed from our solutions. 
We next discuss the
$t^\prime -t^{\prime \prime}-J_z $ model, and then the
$t-J_z$ and
$t-t^{\prime}-t^{\prime \prime} -J_z $ models.
For the $t-J_z$ model, we perturbatively construct
an extended quasi-pair.
As a step towards exploring the
range of validity of our approach, 
we compare with results
for a 2x2 plaquette and numerical studies. 
Lastly, we discuss implications of our results
for the physically relevant parameter regime,
including the question of the sufficiency of the 
$t - t^\prime -t^{\prime \prime}-J$ model for
capturing high-$T_c$ behavior.

Specifically, we consider the Hamiltonian
\begin{equation} 
 H = H_0 + H_1 + H_2 + H_3,
 \label{h} 
\end{equation} 
where
\begin{eqnarray}
 H_0 =
 J_z &\sum_{ x,y }& \{  ( S_{x,y}^{z} S_{x+1,y}^{z} + 
 S_{x,y}^{z} S_{x,y+1}^{z} ) \nonumber\\
 && - {1 \over 4} ( n_{x,y} n_{x+1,y} + n_{x,y} n_{x,y+1} ) \},
 \label{h0}
\end{eqnarray}
\begin{eqnarray}
 H_1 = 
 (-t) &\sum_{x,y,\sigma}& \{
 ( {\tilde c}_{x,y,\sigma}^\dagger 
 {\tilde c}_{x+1,y,\sigma} + H.c.) \nonumber\\
 && + \,
 ( {\tilde c}_{x,y,\sigma}^\dagger 
 {\tilde c}_{x,y+1,\sigma} + H.c.) \},
 \label{h1}
\end{eqnarray}
\begin{eqnarray}
 H_2 = 
 (-t^\prime) &\sum_{x,y,\sigma}& \{
 ( {\tilde c}_{x,y,\sigma}^\dagger 
 {\tilde c}_{x+1,y+1,\sigma} + H.c.) \nonumber\\
 && + \,
 ( {\tilde c}_{x,y,\sigma}^\dagger 
 {\tilde c}_{x+1,y-1,\sigma} + H.c.) \},
 \label{h2}
\end{eqnarray}
and 
\begin{eqnarray}
 H_3 = 
 (-t^{\prime\prime}) &\sum_{x,y,\sigma}& \{
 ( {\tilde c}_{x,y,\sigma}^\dagger 
 {\tilde c}_{x+2,y,\sigma} + H.c.) \nonumber\\
 &&  + \,
 ( {\tilde c}_{x,y,\sigma}^\dagger 
 {\tilde c}_{x,y+2,\sigma} + H.c.) \}.
 \label{h3}
\end{eqnarray}
Here, $x$ and $y$ denote the coordinates of an $L$x$L$ lattice
with periodic boundary conditions
and even $L$, and $\sigma = \pm 1 \,\, (\uparrow, \downarrow)$ refers to
electron spin.
${\tilde c}_{x,y,\sigma} = c_{x,y,\sigma} ( 1 - n_{x,y,-\sigma} )$, 
enforcing the condition of no double occupancy.
$S_{x,y}^{z} = {1 / 2} \,\, ( n_{x,y,\uparrow} - n_{x,y,\downarrow} )$
and 
$n_{x,y} = n_{x,y,\uparrow} + n_{x,y,\downarrow}$.
We do not explicitly 
consider here the spin-flip part of the magnetic interaction
\begin{eqnarray}
 H_{\perp} =
 \Bigl( { {J_{\perp}} \over 2 } \Bigr)
 \sum_{ x,y }  \left\{ ( S_{x,y}^{+} S_{x+1,y}^{-} +
 S_{x,y}^{+} S_{x,y+1}^{-} ) + H.c. \right\},
\label{jperp}
\end{eqnarray}
where $ S_{x,y}^{+} = c^{\dagger}_{x,y,\uparrow} c_{x,y,\downarrow}$
and $ S_{x,y}^{-} = c^{\dagger}_{x,y,\downarrow} c_{x,y,\uparrow}$.
(The full $t - t^{\prime} - t^{\prime \prime} - J$ model is recovered when
$J_{\perp} = J_z$.)

At half filling each site is occupied by exactly  one electron,
and the doubly degenerate ground state of $H_0$
is then that of a N\'eel
antiferromagnet.
We choose $|\Phi_{a}>$ to denote the state with electron spins
$\sigma(x,y) = (-1)^{x+y}$ and $|\Phi_{b}>$ to denote the state
with $\sigma(x,y) = (-1)^{x+y+1}$.
We define the operator $a_{x,y} = c_{x,y,\sigma(x,y)}$ with
$\sigma(x,y) = (-1)^{x+y}$,
and the operator $b_{x,y} = c_{x,y,\sigma(x,y)}$ with
$\sigma(x,y) = (-1)^{x+y+1}$
Although our calculations and results are independent of the ordering
convention chosen, we will denote for specificity
\begin{equation} 
 |\Phi_a> = (a^{\dagger}_{L,L}...a^{\dagger}_{1,L})...
  (a^{\dagger}_{L,2}...a^{\dagger}_{1,2})
  (a^{\dagger}_{L,1}...a^{\dagger}_{1,1})|0>,
 \label{gsa}
\end{equation}
with an analogous definition for $|\Phi_b>$.

We now dope the half-filled 
state $|\Phi_a>$
with two holes and consider the strong-coupling
limit ($J_z >> |t|, |t^\prime|, |t^{\prime \prime}|$). In this limit,
there will be an energy cost of order $J_z$ if the two holes are
{\it not} nearest neighbors (n.n.). Hence, to zeroth order,
the (highly degenerate) two-hole ground state is spanned by the set of all 
n.n. hole pairs.
We denote the state with a horizontal n.n. hole pair at sites
$(x,y)$ and $(x+1,y)$ as
\begin{equation} 
 |h_{x,y}> = a_{x+1,y} a_{x,y} |\Phi_a>,
 \label{hxy}
\end{equation}
and the state with a vertical n.n. hole pair at sites $(x,y)$ 
and $(x,y+1)$ as
\begin{equation} 
 |v_{x,y}> = a_{x,y+1} a_{x,y} |\Phi_a>.
 \label{vxy} 
\end{equation} 
The $|h_{x,y}>$'s and $|v_{x,y}>$'s provide a complete, orthonormal basis
for the two-hole ground state of $H_0$.

It costs an energy of order $J_z$ if one of the n.n. holes
hops to a n.n. site through the hybridization matrix element $t$.
However, there is no energy cost
for hops corresponding to $t^{\prime}$ or $t^{\prime \prime}$, as
long as the two holes remain nearest neighbors after the hop.
Thus, to lowest order in $1 / J_z$, it is only necessary
to diagonalize the Hamiltonian $H_2 + H_3$ in the
subspace spanned by the $|h_{x,y}>$'s and $|v_{x,y}>$'s; i.e., it is only 
necessary to consider the
$t^\prime -t^{\prime \prime}-J_z $ model.
We note that in this
limit the $t^\prime -t^{\prime \prime}-J_z $ model becomes
isomorphic to the strong-coupling limit of the
antiferromagnetic van Hove model of \cite{afvh}.

We consider first the $t^\prime -J_z $ model, involving only the
$H_2$ (diagonal) hopping term.
Defining
\begin{equation} 
 |h_{k_{x},k_{y}}> = {1 \over L} 
 \sum_{ x,y } e^{-{ { 2 \pi i k_{x} x } \over L }}
 e^{-{ { 2 \pi i k_{y} y } \over L }} \, |h_{x,y}>
\label{hk} 
\end{equation} 
and
\begin{equation} 
 |v_{k_{x},k_{y}}> = {1 \over L} 
 \sum_{ x,y } e^{-{ { 2 \pi i k_{x} x } \over L }}  
 e^{-{ { 2 \pi i k_{y} y } \over L }} \, |v_{x,y}>,
\label{vk} 
\end{equation} 
with $k_{x},k_{y} = 0,1... \, L-1$,
we obtain the lowest order wave functions
\begin{eqnarray} 
 |\psi^{\pm}_{k_{x},k_{y}}> &=&
 {1 \over {\sqrt 2}} \,
 \Bigl\{ e^{-{ { \pi i k_{x} x } \over L }}
 \, |h_{k_{x},k_{y}}> \nonumber\\
 && \, \pm \,\,  {\rm sgn}(t^{\prime}) \,
 e^{-{ { \pi i k_{y} y } \over L }}  \, |v_{k_{x},k_{y}}> \, \Bigr\}
\label{tpwf}  
\end{eqnarray}  
with energies
\begin{equation}  
 \epsilon^{\pm}_{k_{x},k_{y}} = \pm \, 4 \, |t^{\prime}| 
  \> \sin \Bigl( { { \pi k_{x} } \over L } \Bigr)  
  \> \sin \Bigl( { { \pi k_{y} } \over L } \Bigr).
\label{tpe}   
\end{equation}   
 
Since
$0 \le \sin (\pi k_{x} / L), \sin (\pi k_{y} / L) \le 1$,
the minus sign gives the branch of lower energy.
The lowest energy state $| \psi ^{(a)} _{0} >$,
with energy $-4 \, |t^{\prime}|$, occurs when
$k_{x} = L / 2$ and $k_{y} = L / 2$
(i.e., $( \pi, \pi )$).
Rewriting in terms of the $a_{x,y}$'s and neglecting overall
phase factors, one obtains
\begin{eqnarray}
| \psi ^{(a)} _{0} > = {1 \over {L \sqrt 2}}
&\sum_{x,y}& (-1)^{x+y} \{ a_{x+1,y} a_{x,y} \nonumber\\
&& \, - \,\, {\rm sgn}(t^{\prime})\,  a_{x,y+1} a_{x,y} \} | \Phi_{a}>.
\label{tpgs}
\end{eqnarray}
When $t^{\prime} > 0$ (sgn($t^{\prime}$) = 1),
the sum over hole pair operators in Eq. \ref{tpgs}
changes sign upon a 90 degree
rotation around a lattice point, giving the pair $d$-wave symmetry
(specifically, $d_{x^2 - y^2}$\cite{annett,scalrev}).
When $t^{\prime} < 0$, there are no such sign changes,
giving $s$-wave symmetry
(specifically, extended-$s$\cite{annett,scalrev}).

If one adds to Eq. \ref{tpgs}
the appropriately-phased pair operator 
for two holes doped into the
ground state $| \Phi_{b} >$,
one obtains for $t^{\prime} > 0$ the usual (unnormalized)
n.n. singlet $d_{x^2 - y^2}$
pair operator
\begin{eqnarray}
{1 \over L}
&\sum_{x,y}&
\Bigl\{ \bigl( c_{x,y,\uparrow} c_{x+1,y,\downarrow} -  
               c_{x,y,\downarrow} c_{x+1,y,\uparrow} \bigr) \nonumber\\
   && - \bigl( c_{x,y,\uparrow} c_{x,y+1,\downarrow} -  
               c_{x,y,\downarrow} c_{x,y+1,\uparrow} \bigr) \Bigr\} ,
\label{ccd}
\end{eqnarray}
with $t^{\prime} < 0$ giving the analogous singlet extended-$s$ operator.
With different relative phases, one can also obtain $m=0$ triplet pairs;
because quantum spin fluctuations are not included in the $t-J_z$
model, the two cases cannot be differentiated at this
level.

\begin{figure}[h]
\centering
\mbox{}
\psfig{figure=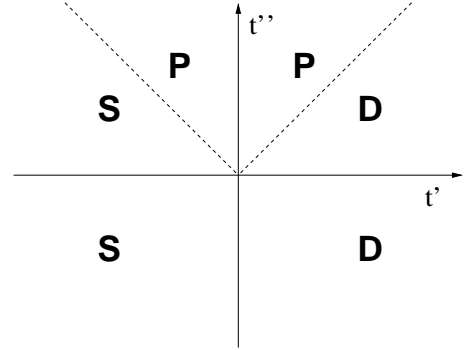,width=6cm,angle=270}
\vspace{0.1cm}
\caption{Hole pair symmetry in the strong-coupling limit of the
$t^\prime - t^{\prime\prime}-J_z$ model
as a function of $t^\prime$ and $t^{\prime\prime}$.
Here, ``$D$'' denotes $d_{x^{2} - y^{2}}$, ``$P$'' denotes
$p_x$ or $p_y$, and ``$S$'' denotes extended-$s$.}
\label{fig1}
\end{figure}

For the more general
$t^\prime -t^{\prime \prime}-J_z $ model,
one obtains the (unnormalized) wave functions
\begin{eqnarray} 
 |\psi^{\pm}_{k_{x},k_{y}}&>& \, = \,
 e^{-{ { \pi i k_{x} x } \over L }}  
 (4 t^{\prime}) s_{x} s_{y} |h_{k_{x},k_{y}}> \nonumber\\
 && \, + \, e^{-{ { \pi i k_{y} y } \over L }}
 \Bigl[ (2 t^{\prime \prime})
 (s_{y}^{2} -  s_{x}^{2}) \pm \tau_{x,y} \Bigr]
  |v_{k_{x},k_{y}}>
\label{ttwf1}  
\end{eqnarray}  
with energies
\begin{eqnarray} 
 \epsilon^{\pm}_{k_{x},k_{y}} = 
 (-2 t^{\prime \prime}) (1 - s_{x}^{2} -  s_{y}^{2}) \pm \tau_{x,y} \, ,
\label{tte}
\end{eqnarray}  
where $s_{x} = \sin ( \pi k_{x}  / L )$, 
$s_{y} = \sin ( \pi k_{y} /  L )$, and
\begin{eqnarray}
\tau_{x,y} = 2 \Bigl\{ ( t^{\prime \prime} )^{2}
 ( s_{x}^{2} - s_{y}^{2} )^{2}
+ 4 ( t^{\prime} )^{2} s_{x}^{2} s_{y}^{2} \Bigr\} ^{ 1 \over 2}.
\label{tau}
\end{eqnarray}
As a function of $t^{\prime}$ and $t^{\prime \prime}$, we find that
the ground state symmetry of the pair is as shown in Fig. 1. The
$p$-wave pair operators can be either $p_x$
\begin{eqnarray}
{1 \over L}
 \sum_{ x,y } e^{-{ { 2 \pi i k_{y} y } \over L }}
a_{x,y} ( a_{x+1,y} - a_{x-1,y} )
\label{px}
\end{eqnarray}
or $p_y$
\begin{eqnarray}
{1 \over L}
 \sum_{ x,y } e^{-{ { 2 \pi i k_{x} x } \over L }}
a_{x,y} ( a_{x,y+1} - a_{x,y-1} ).
\label{py}
\end{eqnarray}
The $p_x$ states have energies independent of $k_y$, and the
$p_y$ states have energies independent of $k_x$.
Both $p$-wave pair operators change sign under a 180 degree rotation.

We next consider the strong-coupling limit of the $t-J_z$ model.
To lowest order, we find that
this maps onto the above strong-coupling limit
of the 
$t^{\prime} - t^{\prime \prime} - J_z$ model with 
\begin{eqnarray}
 t_{eff.}^{\prime} = t_{eff.}^{\prime \prime} = 
{ 2 \over 3 }
 \left( { { t^{2} } \over { J_{z} } } \right).
\label{teff}
\end{eqnarray}
From Eq. \ref{tte}, 
the lower band of the pair excitation spectrum then
becomes flat, with wave functions
\begin{eqnarray} 
 |\psi^{-}_{k_{x},k_{y}}> &=&
 {1 \over {\sqrt 2}} \,
 \Bigl\{ e^{-{ { \pi i k_{x} x } \over L }}
 \, s_{y} |h_{k_{x},k_{y}}> \nonumber\\
 && \, \, \, \, \, \, \, \,  - \,  \,
 e^{-{ { \pi i k_{y} y } \over L }}  \,
 s_{x} |v_{k_{x},k_{y}}> \, \Bigr\}.
\label{twf}  
\end{eqnarray}  
Flat bands were
also found \cite{shraiman,trugman}
for related models and/or treatments.
In \cite{cumulant},
a five-fold degeneracy of strong-coupling
$t-J_z$ pairs of
$d$ or $p$ symmetry was noted.

We see from Eq. \ref{teff}
that, to lowest order, the strong-coupling
$t-J_z$ model lies on the (rightmost)
boundary in Fig. 1 between $d$-wave and $p$-wave symmetry.
In the next higher order, neglecting constant additive terms,
the energies of the lower band separate into
\begin{eqnarray}
 \epsilon^{-}_{k_{x},k_{y}} &=&
 \Bigl( - { 8 \over 45 } \Bigr)
 \Bigl( { { t^{4} } \over { J_{z}^3 } } \Bigr)
 \Bigl( 2 - c_x - c_y \Bigr)^{-1} \nonumber\\
 && \bigl\{ c_{x}^{2} + c_{y}^{2} + 4 c_{x} c_{y}
     - 31 c_{x} - 31 c_{y} + 56 \bigr\},
\label{fourth}
\end{eqnarray}
where here $c_{x} = \cos ( 2 \pi k_{x} / L )$
and
$c_{y} = \cos ( 2 \pi k_{y} / L )$.
We then find (in agreement with \cite{cumulant})
that the pure $d$-wave ($t^{\prime} > 0$)
state of Eq. \ref{tpgs}
is selected as the ground state.
However, the closeness to $p$-wave symmetry may 
provide an explanation for
the low-energy $p$-wave ``quasi-pair'' peaks seen numerically
in small $t-J$ and $t-J_z$ clusters\cite{dagottop}.
Because of this similar $t-J$ and $t-J_z$ behavior,
referring to Fig. 1 and assuming ground state pairs of
pure symmetry, we predict that
adding the appropriate $t^{\prime}$ and/or
$t^{\prime \prime}$ 
to the $t-J_z$ or $t-J$
models with
$J_{z} / t$ or $J / t$ sufficiently large
should drive the models to $p$-wave hole pair symmetry,
and perhaps
even $p$-wave superconductivity. (In one dimension,
a n.n.n. $t^{\prime} > 0$ will also give $p$-wave
hole-pair symmetry in the $J_{z} >> |t|,|t^{\prime}|$
limit.)

One can
perturbatively construct increasingly extended quasi-pair states 
for the $t-J_z$ model.
Combining results for the n.n. $d$-wave pair operators
for ground states $| \Phi_a>$ and $| \Phi_b>$,
one finds the lowest order correction for the singlet pair
operator of Eq. \ref{ccd}
\begin{eqnarray}
\Bigl( - {4 \over 3} \Bigr)
\Bigl( {t \over {J_z}} \Bigr)
\Bigl( {1 \over L} \Bigr)
\sum_{x,y,\sigma}
\sigma  & \nonumber\\
 \Bigl\{ \bigl( c^{\dagger}_{x+1,y,\sigma} c_{x+1,y,-\sigma} -  
 & c^{\dagger}_{x,y+1,-\sigma} c_{x,y+1,\sigma} \bigr) \nonumber\\
          c_{x+1,y+1,\sigma} c_{x,y,\sigma} \,\,\,\,\, & \nonumber\\
  \mbox{} + \bigl( c^{\dagger}_{x+1,y,\sigma} c_{x+1,y,-\sigma} -  
 & c^{\dagger}_{x,y-1,-\sigma} c_{x,y-1,\sigma} \bigr) \nonumber\\
          c_{x+1,y-1,\sigma} c_{x,y,\sigma} \Big\}. &
\label{dpert}
\end{eqnarray}
When operating on the appropriate N\'eel state,
each of the above terms consists of a diagonal hole pair
``dressed'' with a singlet pair of electrons straddling the bond
connecting the pair of holes, as was recently found in numerical
$t-J$ simulations\cite{whitescal}.
We note that the contribution
from pairs a distance of
two lattice sites apart, nominally also of
order $t / J_z$,
vanishes identically
in this order.
This may provide an explanation for why only
n.n. and diagonal
hole correlations appear to dominate in the $t-J$
model near half filling for moderate to
large $J / t$\cite{whitescal,didier}.

If one adds the necessary terms to the operator of Eq. \ref{dpert}
to impose rotational invariance, one obtains
the composite pair operator invented in \cite{didier}
to give a diagonal singlet pair
with $d_{x^2 - y^2}$ symmetry.
The non-invariant 
operator of Eq. \ref{dpert}, which emerges naturally from
perturbation theory, also has 
$d_{x^2 - y^2}$ symmetry.

We also note that, since we calculate energy spectra and
wave functions, our results and approach can be used to
calculate finite-temperature and real frequency
properties.
However, we do not pursue that here.

\begin{figure}[h]
\centering
\mbox{}
\psfig{figure=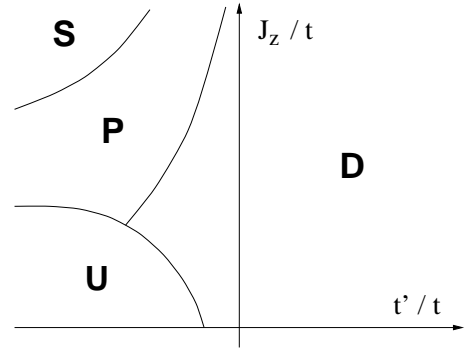,width=6cm,angle=270}
\vspace{0.1cm}
\caption{Qualitative diagram of predicted hole pair symmetry
for the $t - t^\prime-J_z$ model.
``$D$'', ``$P$'',
and ``$S$'' denote the same as in Fig. 1, and
no prediction is made for region ``$U$''.}
\label{fig2}
\end{figure}

As one step towards investigating the range of validity
of our approach, we performed analytic calculations of the
$t-t^{\prime}-J_{z}-J_{\perp}$ model
(see Eq. \ref{jperp})
on a 2x2 plaquette.
In general, we found that ground states
remained smoothly connected
as $J_z$ was reduced from strong coupling and also as $J_{\perp}$
was turned on.
$t-J$ numerical results \cite{cumulant,didier}
also indicate that 
features of the strong-coupling limit may persist down to
intermediate coupling ($J/t \approx 0.4 - 0.5$).
Together, the above support the strong-coupling 
$t- t^\prime -t^{\prime \prime}-J_z $ model as a useful starting
point for exploring
the intermediate-coupling
$t- t^\prime -t^{\prime \prime}-J$ model.
Based on this and our strong-coupling results
(and, again, assuming pairs of pure symmetry),
we show in Fig. 2 qualitative predictions of the hole pair
symmetry for the $t - t^{\prime} - J_z$ model.
We believe these predictions apply to the
$t - t^{\prime} - J$ model as well,
with a comparatively smaller $p$-wave region due to
larger energy differences between $t-J$ $p$-wave and $d$-wave
pair states \cite{dagottop}. An additional
$t^{\prime \prime} > 0$ would enlarge the $p$-wave region.

Reductions from CuO$_2$ three-band models \cite{param}, as well as
comparison with ARPES results for a single doped hole\cite{onefit},
suggest that $| t^{\prime} / t|$ and $| t^{\prime \prime} / t|$ may
be substantial.
$t > 0$, and estimates
for $t^{\prime}$ and $t^{\prime \prime}$
are typically in the ranges
$t^{\prime} \approx (-0.1)t - (-0.5)t$, and
$t^{\prime \prime} \approx 0.0 - (0.3)t$.
Both these signs of $t^{\prime}$ and $t^{\prime \prime}$
could tend to drive the pairing symmetry to $p$-wave,
raising the issue of the hole pair
symmetry in the intermediate-coupling regime.
($s$-wave is also possible, though we believe it less likely
at intermediate coupling.)
It would be interesting to numerically
explore whether the symmetry of two doped holes in the
$t- t^\prime -t^{\prime \prime}-J$ model is in fact $d$-wave for
the experimentally relevant values of
$t, t^\prime, t^{\prime \prime}$, and $J$ (e.g., $J/t \approx 0.4$). 
Drawing conclusions from exact diagonalization may be challenging 
due to finite-size effects: either $s$-wave or $p$-wave
hole pair symmetries were found
on lattices of sizes 16, 18, and 20 for the
$t - t^{\prime} - J$ model with realistic parameters\cite{onefitsps}.
Another possible tool is higher order numerical 
ground state perturbation theory \cite{cumulant}.
If the symmetry
were established to be $p$-wave rather than $d$-wave,
it would suggest that the 
$t- t^\prime -t^{\prime \prime}-J$ model
by itself could be incomplete
as a model for high-$T_c$ superconductivity.
In that case,
one possibility for restoring $d$-wave symmetry
could be the addition of electron-phonon coupling in the
$d$-channel \cite{phonon}.
In either case,
it may also be of interest to explore whether the existence
of or nearness to $p$-wave symmetry, which effectively reduces the
dimensionality of the hole pair wave function from 2D to 1D,
might play a role in the ``striping'' recently observed in certain
of the high-$T_c$ cuprates \cite{striping}.

In summary, we have investigated analytically
the pair symmetry and excitation
spectra of two holes doped into the half-filled
$t- t^\prime -t^{\prime \prime}-J_z $ model
in the strong-coupling
limit.
In lowest order, this reduces to considering the 
$t^{\prime}- t^{\prime \prime} -J_z $ model, where we found
regions of $d$-wave, $s$-wave, and $p$-wave symmetry.
We next found that the $t-J_z$ model in lowest order was
on the boundary between $d$-wave and $p$-wave pair symmetry,
with a flat lower pair excitation spectrum.
In higher order, $d$-wave pairing was selected.
However, because of the closeness to $p$-wave symmetry,
we predict 
that the appropriate 
$t^\prime$ and/or $t^{\prime \prime}$
added to the $t-J_z$ or $t-J$
models with intermediate to large $J_z$ or $J$
should drive them into $p$-wave pairing,  and perhaps
even $p$-wave superconductivity.
We constructed a perturbative correction to the
nearest neighbor $d$-wave pair,
and compared with the $d$-wave 
composite operator invented in \cite{didier}. We
explored ranges of validity of this perturbative approach using
a 2x2 plaquette and results from other work \cite{cumulant,didier}.
Lastly, we discussed implications for
the experimentally relevant parameter regime. These included
the possibility of $p$-wave symmetry for two doped holes, which would
suggest that the
$t - t^{\prime}- t^{\prime \prime} -J$ model could be incomplete as
a high-$T_c$ model, and the possible relevance of our results
to ``striping''.


\begin{references}

\bibitem{dexp1} D.J. Van Harlingen,
Rev. Mod Phys. {\bf 67}, 515 (1995). 

\bibitem{annett} J.F. Annett, N. Goldenfield, and A.J. Leggett,
in {\bf Physical Properties of High Temperature
Superconductors}, Vol. V, ed. D.M. Ginsberg (1996).

\bibitem{dagottorev} E. Dagotto,
Rev. Mod Phys. {\bf 66}, 763 (1994), and refs. therein.

\bibitem{schrieffer} J.R. Schrieffer,
Solid State Commun. {\bf 92}, 129 (1994).

\bibitem{reviews} A.P. Kampf, Phys. Rep. {\bf 249},
219 (1994); W. Brenig, Phys. Rep. {\bf 251}, 155 (1995);
and refs. therein. 

\bibitem{scalrev} D.J. Scalapino,
Phys. Rep. {\bf 250}, 329 (1995).

\bibitem{scaltrug} D.J. Scalapino and S.A. Trugman,
J. Phil. Mag. B {\bf 74}, 607 (1996).

\bibitem{whitescal} S.R. White and D.J. Scalapino,
Phys. Rev. B {\bf 55}, 6504 (1997).

\bibitem{pseudo1} R.S. Markiewicz, cond-mat/9611238,
and refs. therein.

\bibitem{pseudo2} A.V. Puchkov et al.,
cond-mat/9611083.

\bibitem{preform} J.M. Harris et al.; cond-mat/9611010, 9705304.

\bibitem{nan} N. Trivedi and M. Randeria,
Phys. Rev. Lett. {\bf 75}, 312 (1995).

\bibitem{dagottop} E. Dagotto et al.,
Phys. Rev. B {\bf 42}, 2347 (1990).

\bibitem{tjz}
Z. Liu and E. Manousakis,
Phys. Rev. B {\bf 45}, 2425 (1992);
J. Gan and P. Hedeg\aa{}rd,
Phys. Rev. B {\bf 53}, 911 (1996);
O.A. Starykh and G.F. Reiter,
Phys. Rev. B {\bf 53}, 2517 (1996).

\bibitem{cumulant} P. Prelov\u sek et al.,
Phys. Rev. B {\bf 42}, 10706 (1990).

\bibitem{afvh} E. Dagotto et al.,
Phys. Rev. Lett. {\bf 74}, 728 (1995).

\bibitem{shraiman} B.I. Shraiman and E.D. Siggia,
Phys. Rev. Lett. {\bf 60}, 740 (1988).

\bibitem{trugman} S.A. Trugman,
Phys. Rev. B {\bf 37}, 1597 (1988).

\bibitem{didier} D. Poilblanc,
Phys. Rev. B {\bf 49}, 1477 (1994).

\bibitem{param}
H. Eskes et al., Physica C {\bf 160}, 1989;
M. Hybertsen et al.,
Phys. Rev. B {\bf 41}, 11068 (1990);
T. Tohyama and S. Maekawa, J. Phys. Soc. Jpn. {\bf 59}, 1760 (1990);
D.C.  Mattis and J.M.  Wheatley,
Mod.  Phys.  Lett.  {\bf 9}, {1107} (1995);
V.I. Belinicher et al.,
Phys. Rev. B {\bf 53}, 335 (1996);
L.F. Feiner et al.,
Phys. Rev. B {\bf 53}, 8751 (1996);
R. Hayn et al., cond-mat/9606043.

\bibitem{onefit}
A. Nazarenko et al.,
Phys. Rev. B {\bf 51}, 8676 (1995);
P.W. Leung and R.J. Gooding,
Phys. Rev. B {\bf 52}, 15711 (1995);
B.O. Wells et al.,
Phys. Rev. Lett. {\bf 74}, 964 (1995);
L.F. Feiner et al.,
Phys. Rev. Lett. {\bf 76}, 4939 (1996);
T. Xiang and J.M. Wheatley, 
Phys. Rev. B {\bf 54}, 12653 (1996);
V.I. Belinicher et al.,
Phys. Rev. B {\bf 54}, 14914 (1996);
D. Duffy et al., cond-mat/9701083.

\bibitem{onefitsps} R. Eder, Y. Ohta, and
G.A. Sawatsky,
Phys. Rev. B {\bf 55}, 3414 (1997).

\bibitem{phonon} O.K. Andersen et al., cond-mat/9703238.

\bibitem{striping}
Bianconi et al.,
Phys. Rev. Lett. {\bf 76}, 3412 (1996);
Tranquada et al.,
Phys. Rev. Lett. {\bf 78}, 338 (1997); and
refs. therein.

\end{references}
\end{document}